\documentclass[apj]{emulateapj}
\slugcomment{To appear in ApJ}

\def\lsi{\object{LS~I~+61\,303\,}}
\def\degs{\ifmmode ^{\circ}\else$^{\circ}$\fi}
\def\asec{\ifmmode ^{\prime\prime}\else$^{\prime\prime}$\fi}
\def\msun{~M_{\odot}}

\def\degs{\ifmmode ^{\circ}\else$^{\circ}$\fi}
\def\ergs{\mbox{~ergs~s$^{-1}$}}
\def\ergcmcms{\mbox{~ergs~cm$^{-2}$~s$^{-1}$}}
\def\EE#1{\times 10^{#1}}
\def\kms{\mbox{\,km~s$^{-1}$}}
\def\kms{\mbox{\,km s$^{-1}$}}
\def\lsim{\!\!\!\phantom{\le}\smash{\buildrel{}\over
{\lower2.5dd\hbox{$\buildrel{\lower2dd\hbox{$\displaystyle<$}}\over
                                \sim$}}}\,\,}
\def\gsim{\!\!\!\phantom{\ge}\smash{\buildrel{}\over
{\lower2.5dd\hbox{$\buildrel{\lower2dd\hbox{$\displaystyle>$}}\over
                              \sim$}}}\,\,}

\begin{document}
\title{Multi-wavelength (radio, X-ray and $\gamma$-ray) observations of the $\gamma$-ray binary \lsi
\altaffilmark{1}}
\shorttitle{Multi-wavelength observations of \lsi}
\shortauthors{Albert et al.}

%
\author{
J.~Albert\altaffilmark{a},
E.~Aliu\altaffilmark{b},
H.~Anderhub\altaffilmark{c},
P.~Antoranz\altaffilmark{d},
M.~Backes\altaffilmark{e},
C.~Baixeras\altaffilmark{f},
J.~A.~Barrio\altaffilmark{d},
H.~Bartko\altaffilmark{g},
D.~Bastieri\altaffilmark{h},
J.~K.~Becker\altaffilmark{e}
% ...
\\[0.2cm]
(The MAGIC Collaboration)
\\[0.2cm]
M.A.~Guerrero\altaffilmark{t}, 
A.~Alberdi\altaffilmark{t}, 
Z.~Paragi\altaffilmark{y},
T.W.B.~Muxlow\altaffilmark{z}, 
P.~Diamond\altaffilmark{z}
}

% ...
\altaffiltext{a} {Universit\"at W\"urzburg, D-97074 W\"urzburg, Germany}
\altaffiltext{b} {IFAE, Edifici Cn., Campus UAB, E-08193 Bellaterra, Spain}
\altaffiltext{c} {ETH Zurich, CH-8093 Switzerland}
\altaffiltext{d} {Universidad Complutense, E-28040 Madrid, Spain}
\altaffiltext{e} {Universit\"at Dortmund, D-44227 Dortmund, Germany}
\altaffiltext{f} {Universitat Aut\`onoma de Barcelona, E-08193 Bellaterra, Spain}
\altaffiltext{g} {Max-Planck-Institut f\"ur Physik, D-80805 M\"unchen, Germany}
\altaffiltext{h} {Universit\`a di Padova and INFN, I-35131 Padova, Italy}
% ...
\altaffiltext{t} {Instituto de Astrof\'\i sica de Andaluc\'\i a - CSIC, E-18008 Granada, Spain}
% ...
\altaffiltext{y} {Joint Institute for VLBI in Europe, Dwingeloo, Netherlands}
\altaffiltext{z} {JBO, Univ. of Manchester, Macclesfield, UK}
% ...

\begin{abstract}  
We present the results of the first multiwavelength observing campaign
on the high-mass X-ray binary \lsi\ comprising observations at the TeV
regime with the MAGIC telescope, along with X-ray observations with
{\it Chandra}, and radio interferometric observations with the MERLIN,
EVN and VLBA arrays, in October and November 2006. From our MERLIN
observations, we can exclude the existence of large scale ($\sim 100$
mas) persistent radio-jets. Our 5.0 GHz VLBA observations display
morphological similarities to previous 8.4 GHz VLBA observations
carried out at the same orbital phase, suggesting a high level of
periodicity and stability of the processes behind the radio
emission. This makes it unlikely that variability of the radio
emission is due to the interaction of an outflow with variable wind
clumps. If the radio emission is produced by a milliarcsecond scale
jet, it should also show a stable, periodic behavior. It is then
difficult to reconcile the absence of a large scale jet ($\sim$100
mas) in our observations with the evidence of a persistent
relativistic jet reported previously. We find a possible hint of
temporal correlation between the X-ray and TeV emissions and evidence
for radio/TeV non-correlation, which points to the existence of one
population of particles producing the radio emission and a different
one producing the X-ray and TeV emissions. Finally, we present a
quasi-simultaneous energy spectrum including radio, X-ray and TeV
bands.
\end{abstract}


\section{Introduction}  

\lsi\ is a high-mass X-ray binary consisting of a low-mass [$M\sim
(1-4)\,\msun$] compact object orbiting around an early type B0\,Ve
star along an eccentric ($e=0.7$) orbit
\citep[and references therein]{casares05}. The modulation of both, the
radio \citep{gre78,gre02} and X-ray \citep{tay96,paredes97} emissions,
display a period of $P_\textrm{orb} =26.496$~d, attributed to the
orbital motion. The position of the maximum of the radio emission along
the orbit, as well as its intensity, are modulated with a superorbital
period of $P_\textrm{sup} = 1667 \pm 8$ days \citep{gre02}.
\lsi\ is positionally coincident with an EGRET $\gamma$-ray source
\citep{kni97}. Moreover, variable emission at TeV energies has been
recently detected with the MAGIC telescope \citep{alb06}. These
authors found that the peak flux at TeV energies occurs at orbital
phase $\phi_\textrm{orb} \approx 0.65$, while no high-energy emission is detected
around periastron passage ($\phi_\textrm{orb}
\approx 0.23$). From $\sim$50 mas resolution radio images of \lsi\ obtained
with MERLIN, \cite{mas04} suggested the existence of a precessing
relativistic ($\beta\approx0.6$) jet up to angular scales of $\sim$0.1
arcsec, which led them to interpret \lsi\ within the framework of the
microquasar scenario
\citep{valenti06,gustavo07}. However, recent VLBA imaging obtained by
\cite{dha06} over a full orbit of \lsi\ has shown the radio emission to
come from angular scales smaller than about 7~mas (projected size 14
AU at an assumed distance of 2~kpc). This radio emission
appeared cometary-like, interpreted to be pointing away from the high
mass star within a particular scenario,
and no relativistic motion, nor halos, nor larger-scale structures
were detected at any phase of the orbit. Based on these findings,
\cite{dha06} concluded that the radio and TeV emissions from
\lsi\ are originated by the
interaction of the wind of a young pulsar with that of the stellar
companion \citep{maraschi81,dub06}.

In this article, we present and discuss the results of a
multi-wavelength campaign including radio, X-ray, and TeV $\gamma$-ray
observations on \lsi, aimed at shedding light on the physical
processes going on in the system, as well as yielding useful input for
a detailed, time-dependent modeling of this relevant system. For this,
we study the correlations between the different detected emissions in
terms of their morphological, temporal and spectral features, both in
the intra-day and day-to-day timescales.

\section{Multiwavelength observations} 

Table \ref{tbl-log} and Figure~\ref{fig:obs} summarize our
observations of \lsi, carried out during October and November 2006.
In particular, we set up a simultaneous, multiwavelength campaign on
\lsi\ for the 25 and 26 October 2006 using the MERLIN, EVN, and VLBA
interferometers (radio), and the TNG (infra-red), {\it Chandra} (X-rays),
and MAGIC TeV $\gamma$-rays) telescopes.  Unfortunately, bad
weather conditions did not allow us to get any infra-red data, nor did
we get useful data from the MAGIC telescope coincident in time
with simultaneously scheduled {\it Chandra} observations. The period
16--20 November included only MAGIC and MERLIN observations.

The range of observed orbital phase in the October campaign was
$\phi_\textrm{orb}$ = 0.6--0.7, for which the TeV maximum was detected
in the past \citep{alb06}, and $\phi_\textrm{orb}$ = 0.44--0.57 in
the November campaign. The super-orbital phase for both campaigns was
$\phi_\textrm{sup} = 0.4$.

\begin{deluxetable}{lcccc}
\tablewidth{0pt}
\tablecaption{Observing log for the multi-wavelength campaign\label{tbl-log}}
\tablehead{
\colhead{Telescope} & \colhead{Date} &  \colhead{UT range} &
\colhead{$\phi_\textrm{orb}$\tablenotemark{a}} & \multicolumn{1}{l}{Average flux\tablenotemark{b}}
}
\startdata
MERLIN     & 26 Oct & 22:32 -- 04:30 & 0.66 & 34.84$\pm$0.23 \\ 
          & 27 Oct & 10:30 -- 23:59 & 0.68 & 33.74$\pm$0.19 \\ 
          & 28 Oct & 00:00 -- 10:00 & 0.70 & 29.63$\pm$0.18 \\
          & 16 Nov & 14:43 -- 23:59 & 0.44 & 71.41$\pm$0.15 \\
          & 17 Nov & 00:00 -- 23:59 & 0.47 & 70.78$\pm$0.14 \\ 
          & 18 Nov & 00:00 -- 23:59 & 0.50 & 78.98$\pm$0.21 \\
          & 19 Nov & 00:00 -- 13:25 & 0.53 & 69.05$\pm$0.15 \\
          & 20 Nov & 00:00 -- 11:24 & 0.57 & 47.14$\pm$0.25 \\
EVN        & 26 Oct & 22:30 -- 04:30 & 0.65 & 33.29$\pm$0.27 \\
VLBA       & 25 Oct & 21:30 -- 02:50 & 0.61 & 39.49$\pm$0.18 \\
          & 26 Oct & 21:30 -- 02:50 & 0.65 & 32.69$\pm$0.14 \\
{\it Chandra} 
          & 25 Oct & 22:13 -- 04:26 & 0.61 &  1.87$\pm$0.16\\
MAGIC      & 27 Oct & 00:50 -- 04:29 & 0.66 & 1.7$\pm$0.4  \\
          & 16 Nov & 21:33 -- 01:04 & 0.45 &  $<$1.19\\
          & 17 Nov & 20:56 -- 01:00 & 0.48 &  $<$0.80\\
          & 18 Nov & 21:00 -- 22:19 & 0.52 &  $<$1.51\\
          & 19 Nov & 21:00 -- 22:00 & 0.56 &  $<$1.31
\enddata
\tablenotetext{a}{
The orbital and super-orbital phases are computed using
$\textrm{MJD}_0 = 43366.275$, $P_\textrm{orb} = 26.4960$
\citep{gre02}. The super-orbital phase is $\phi_\textrm{sup} = 0.4$
for all the observation period.\\ }
\tablenotetext{b}{
The average measured fluxes are in mJy for radio observations (MERLIN, EVN,
VLBA), $10^{-11}$ \ergcmcms\ for the {\it Chandra} X-ray observations (absorbed
power-law), and $10^{-11}$ photons cm$^{-2}$ s$^{-1}$ for the MAGIC VHE
$\gamma$-ray observations. Upper limits are given at the 95$\%$
confidence level, following the prescription by \cite{Rolke2005}. The
data were taken at 5.0~GHz for the radio observations (except for the
MERLIN data in November 2006, which were taken at 6.0 GHz), at 0.5--10
keV by {\it Chandra}, and at 0.3--5.0 TeV by MAGIC.}
\end{deluxetable}

\subsection{Radio Observations}

We observed \lsi\ with several radio interferometric arrays during
25--26 October 2006, including the Multi-Element Radio Linked
Interferometer (MERLIN) in the UK, the European VLBI Network (EVN),
and the Very Long Baseline Array (VLBA) in the USA, all of which
observed at 5.0 GHz. Additionally, we also monitored the flux and
large angular scale structural and flux density variations of \lsi\
using MERLIN at 5.0 GHz on 27 and 28 October 2006, and at 6.0 GHz for
the period from 16 to 20 November 2006.

The MERLIN array included six antennas (Defford, Cambridge, Knockin,
Darnhall, Jodrell Bank (MkII), and Tabley) for most of our
observations, yielding synthesized beams of about 50 to 70~mas
(corresponding to projected linear sizes of 100 to 140 AU). The EVN
observations on 26 October 2006, which were the first ever carried out
for \lsi using the e-VLBI technique (Szomoru 2006), included five
antennas spread over Europe (Cambridge, Jodrell Bank MkII, Medicina,
Torun, and the Westerbork phased array), yielding a synthesized beam
of about 7 mas (or 14 projected AU). The data were directly streamed
to the correlator at JIVE (Joint Institute for VLBI in Europe,
Dwingeloo, The Netherlands) through the Internet, and this was one of
the first science observations that achieved 256~Mbps sustained data
rate with the e-VLBI technique. The VLBA observations on 25 and 26
October 2006 included ten 25-m antennas spread over the USA, yielding
a beam size of 4.6$\times$2.1~mas, corresponding to a projected linear
resolution of $\sim$9.2 AU and $\sim$4.2 AU in right ascension and
declination, respectively.

\begin{figure*} 
\plotone{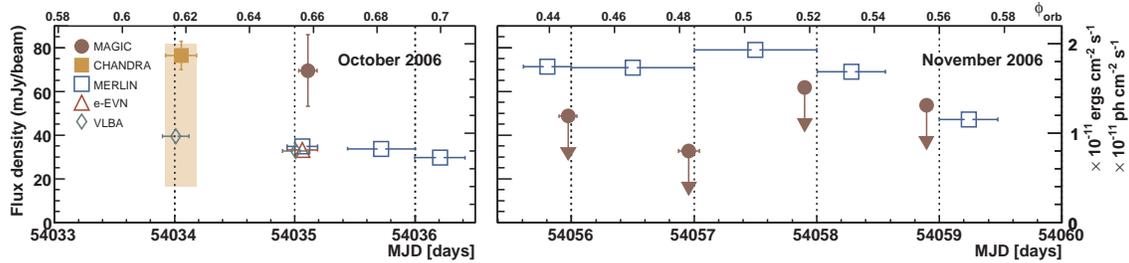}
\caption{Radio, X-ray and VHE $\gamma$-ray light-curves obtained with VLBA,
EVN, MERLIN (left-hand scale), {\it Chandra} (right-hand scale in
$10^{-11}$ \ergcmcms) and MAGIC (right-hand scale in $10^{-11}$
photons cm$^{-2}$ s$^{-1}$) during the two observing periods (October
and November 2006). The horizontal error bars show the time spanned by
the different observations. The shaded area marks the range of X-ray
flux values previously reported (see Table~\ref{tab:xray-hist}). The
upper axis shows the orbital phase using the ephemeris from
\cite{gre02}.}
\label{fig:obs}
\end{figure*}

\begin{figure*}
\plotone{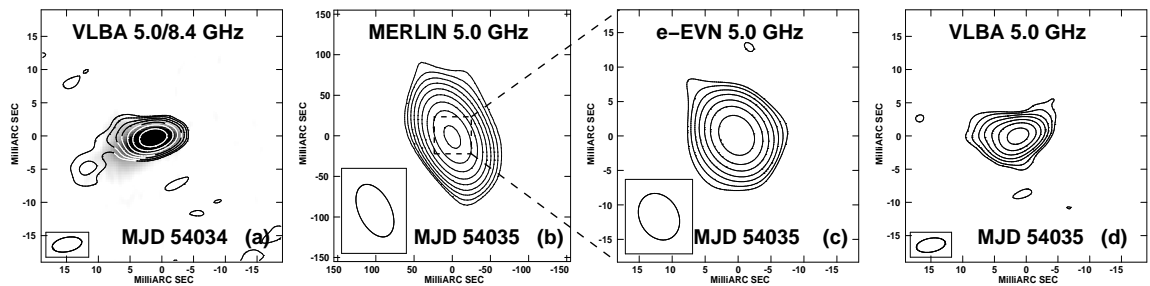}
\caption{Radio images of LS~I~+61\,303\, 
obtained on 25 October 2006 with the VLBA (panel a) and on 26 October
2006 with MERLIN (panel b), EVN (panel c), and VLBA (panel d). The
synthesized beams of the images (bottom left corner in each panel) are
70$\times$41 mas (position angle PA=29$^\circ$), 7.2$\times$5.8 mas
(PA=35$^\circ$), and 4.6$\times$2.1 mas (PA=-81$^\circ$) for the
MERLIN, EVN, and VLBA observations, respectively. The projected linear
resolution of our VLBA observations is thus of 9.2 AU and 4.2 AU in RA
and DEC, respectively. In panel a, we also show overlaid the 8 GHz
VLBA image on 2 February 2006 (grey scale) convolved with our 5 GHz
VLBA beam. This date corresponds to the same phase of LS~I~+61\,303\,
($\phi_\textrm{orb} \approx$0.62), and they show a striking
similarity. In all cases, the origin of coordinates is set at the VLBA
peak of brightness on 25 October 2006, and the contours are drawn at
(3,3\,$\sqrt{3}$,9,...) times the off-source rms (see Table 1). }
\label{fig:radio}
\end{figure*}

\begin{figure*}
\plotone{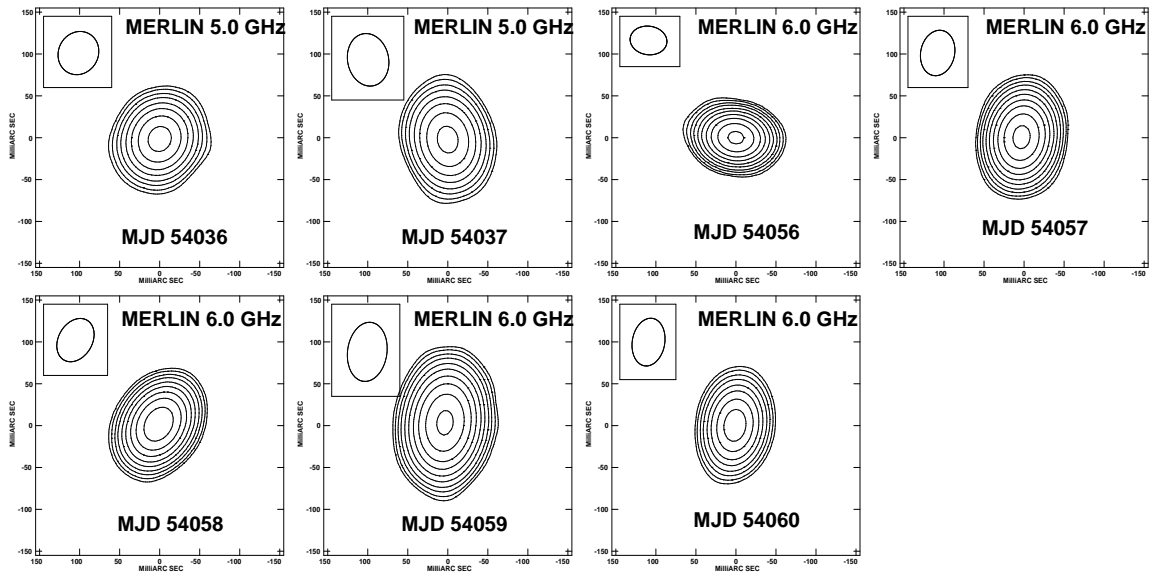}
\caption{Radio images of LS~I~+61\,303\, obtained with MERLIN during
our October 2006 and November 2006 campaigns.  The images correspond,
from left to right and top to bottom, to observations carried out on
27 and 28 October 2006 at a frequency of 5.0 GHz, and on 16, 17, 18,
19, and 20 November 2006 at a frequency of 6.0 GHz, respectively.
Radio contours are drawn at (3,3\,$\sqrt{3}$,9,...) times the
off-source rms (see Table 1). Note that the shape of the radio
brightness distribution of LS~I~+61\,303\, genuinely follows the
synthesized beam for each epoch (upper left inset in each panel).}
\label{fig:merlin}
\end{figure*}

All three interferometric arrays observed \lsi\ in phase-referenced
mode.  \lsi\ and the bright (S$_{\rm 5 GHz }$$\approx 600$mJy), nearby
International Celestial Reference Frame (ICRF) source J0244+6228 were
alternately observed through each observing run, the phases of
J0244+6228 being transfered to the position of \lsi\ in the posterior
data analysis.  J0244+6228 also served as amplitude calibrator for our
observations.  We performed standard calibration and data reduction
within the NRAO Astronomical Imaging Package System ({\it AIPS};
Diamond 1995). We also used standard hybrid mapping techniques within
{\it AIPS} to obtain the flux densities shown in Figure~\ref{fig:obs}
and Table~\ref{tbl-log}, and the radio images shown in Figures
\ref{fig:radio} and \ref{fig:merlin}


\subsection{X-ray Observations}
We obtained {\it Chandra} X-ray observations of \lsi\ on 25 October
2006 through the Director Discretionary Time program
(\dataset [ADS/Sa.CXO#Obs/8273] {Chandra ObsId 8273}).  The
observations were carried out using ACIS-I for a total exposure time
of 20.0 ks.  At the time of the observation, \lsi\ was expected to be
in a high-state and its high X-ray brightness could have resulted in
an excessively high count rate producing appreciable pile-up in the
observations.  In order to minimize pile-up effects, \lsi\ was offset
by 8$\arcmin$ from the ACIS-I aim-point, thus smearing its image and
reducing the count rate at the source peak. We further used 1/4
sub-array, reducing the exposure time of individual frames from the
nominal exposure time of 3.2\,s down to 0.8\,s.

Data reduction of the X-ray observations of \lsi\ have been performed
using the \emph{Chandra} X-ray Center software CIAO V3.4.  The data
reduction included the application of standard filters and rejection
of events with bad grades and those originating from bad pixels.  The
background count rate is consistent with the quiescent background
(Markevitch 2001), and no time intervals of enhanced background needed
to be removed.  The processed observations have a useful exposure time
of 19.1 ks.  Due to the high count rate of \lsi\, out-of-time events
were not negligible and produced a noticeable streak along the
columns.  These have been corrected using the CIAO task
``acisreadcorr''.  Finally, notice that as of CALDB v3.1.0 (June 2005),
the CIAO task ``acis\_process\_events'' is routinely used by the level 1
processing pipeline to mitigate the charge transfer inefficiency (CTI)
that affects the \emph{Chandra} ACIS-I front illuminated chips.
Therefore, no further CTI correction was applied.  Lightcurves and
spectra of \lsi\ were obtained using standard CIAO tasks and analyzed
using HEASARC FTOOLS and XSPEC V11.2.0 routines (Arnaud 1996).

\subsection{VHE $\gamma$-ray Observations}

Observations in the very high energy (VHE) $\gamma$-ray band
($E_\gamma > 100$ GeV) were scheduled with the MAGIC telescope on
26--28 October and 16--19 November 2006 (see
Table~\ref{tbl-log}). These observations are part of an extensive,
standalone observing campaign carried out between September and
December 2006 \citep{alb08}. Bad weather conditions at the
Observatorio del Roque de los Muchachos, however, prevented us from
obtaining useful data on 26 and 28 October 2006.

The observations were carried out in the false-source track (wobble)
mode \citep{Fomin1994}, with two directions at $24^\prime$ distance
and opposite sides of the source direction, which allows for a
reliable estimation of the background with no need of extra
observation time.

The data were analyzed using the standard MAGIC calibration and
analysis software \citep{Albert2006b,Gaug2005}. Data runs with
anomalous event rates were discarded for further analysis. Hillas
variables \citep{Hillas1985} were combined into an adimensional
$\gamma$/hadron discriminator (\emph{hadronness}) and an energy
estimator by means of the Random Forest classification algorithm,
which takes into account the correlation between the different Hillas
variables \citep{Breiman2001,Bock2005}. The incoming direction of the
primary $\gamma$-ray events was estimated using the DISP method,
suited for observations with a single IACT
\citep{Fomin1994,Domingo2005}.


\section{Results}  %
In this section we present the results obtained from the different
multi-wavelength observations we have performed, and put them into the
context of the past measurements in the different bands. They are
summarized in Table~\ref{tbl-log} and Figure~\ref{fig:obs}.

\subsection{Radio Results}

The total radio flux density obtained with MERLIN shows a decline
between October 26 ($\phi_\textrm{orb} = 0.66$) and 28
($\phi_\textrm{orb} = 0.70$), from $\sim 35$ to $\sim 30$ mJy, and a
peak on 18 November ($\phi_\textrm{orb} = 0.50$) at $\sim 80$ mJy. At
the superorbital phase of the observations ($\phi_\textrm{sup} = 0.4$)
the radio source is in the weak state. The predicted flux of the radio
flux is between 50--100 mJy at the orbital phases $\phi_\textrm{orb}
\sim 0.9$. We therefore measure a flux compatible with the predicted one,
although at a much earlier phase value. However, for the weak state,
secondary peaks of comparable flux show up at other orbital phase
values \citep[see, e.g.,][]{dha06}.

In Figure~\ref{fig:radio} we show the radio images corresponding to
25--26 of October, where all three arrays observed simultaneously
\lsi. In Figure \ref{fig:merlin} we show the MERLIN images obtained
on 27 and 28 October 2006 at 5.0 GHz, and on 16 through 20 November at
6.0 GHz. Our MERLIN observations show no evidence for large angular
scale structures, contrary to what had been previously claimed
\citep{mas04}, hence excluding the existence of persistent jets at
these scales. The structure of \lsi\ genuinely follows the beam shape
of the MERLIN array at all epochs. Model
fits to the $u-v$ plane of the MERLIN data show that the projected
intrinsic size of the radio emitting region of \lsi\ is, at any
observing epoch, no larger than $\sim$6~mas ($\sim$12 AU), as
confirmed by our higher resolution images obtained with the EVN and
the VLBA.

Our 5 GHz VLBA images on 25 and 26 October show a very bright and
unresolved component plus an extended radio emission to the
east-southeast (see Figure~\ref{fig:radio}). The general aspect of the
radio emission recalls the structure found by \cite{dha06}. In fact,
panel (a) of Figure~\ref{fig:radio} shows both our 5 GHz VLBA image on
25 October 2006 ($\phi_\textrm{orb} = 0.62$, contours) overlaid on top of the 8.4
GHz VLBA data on 2 February 2006 ($\phi_\textrm{orb} = 0.62$, grey scale), kindly
provided by V.\ Dhawan in advance of publication.

The projected distance from the peak of the radio brightness
distribution to the edge of the extended region is, as imaged with the
VLBA at 5.0 GHz on 26 October 2006, of about $\sim$8.2~mas (16.4~AU).
If the maximum time to fill in this region with radio (synchrotron)
emitting particles is less that the time spanned by our two
consecutive VLBA observations (1 day), the implied outflow velocity is
at least of 21500~km/s, or $v\gsim0.09$~c.

The coordinates of the peak of radio brightness distribution of \lsi,
as obtained from our VLBA observations (using task IMFIT), were
$\textrm{RA}=02^\textrm{h} 40^\textrm{m} 31.6638849^\textrm{s}$,
$\textrm{DEC}=61^\circ 13' 45.592235''$ on October 25 and
$\textrm{RA}=02^\textrm{h} 40^\textrm{m} 31.6638756^\textrm{s}$,
$\textrm{DEC}=61^\circ 13' 45.592496''$ on October 26, with and estimated
accuracy of 12\,$\mu$as and 7\,$\mu$as in RA and DEC, respectively.
This shift of the brightness peak corresponds to a day-to-day
projected speed of 904$\pm$60\kms, in excellent agreement with the
typical value of $\sim$1000\kms\ found by \cite{dha06} along a
complete orbital cycle.

The peak flux density varied significantly between our two VLBA
observing runs, decreasing from 24.3 mJy/b on October 25 to 15.4 mJy/b
on October 26. The total 5~GHz flux density varied from 39.5$\pm$0.2
mJy (radio luminosity $L_{\rm R} = (9.5\pm0.1)\EE{29}$\ergs) to
32.7$\pm$0.2 mJy ($L_{\rm R} = (7.8\pm0.1)\EE{29}$\ergs). We
therefore suggest that the drop seen in the total radio flux density
of \lsi\ between our two consecutive VLBA observations is directly
related to the change in the peak flux density, which hints towards a
physical link with structural variations in the innermost
($\lsim$3~mas projected radius, or 6 AU at a distance of 2~kpc) region
of the source, which our observations cannot resolve.

\subsection{X-ray Results}

\begin{figure}
\epsscale{1.}
\plotone{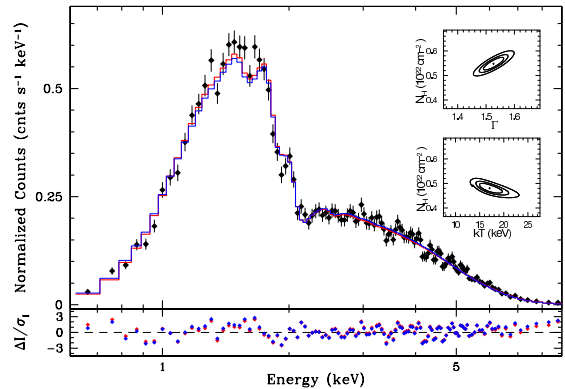}
\caption{ {\it Chandra} ACIS-I spectrum of LS~I~+61\,303 over-plotted
with the best-fit absorbed power law (red histogram) and
Bremsstrahlung (blue histogram) models (see Table~\ref{tab:xray-hist}
for details). The lower panel shows the relative residuals of the fits
($\Delta$I) in terms of the bin standard deviation ($\sigma$) for both
the absorbed power law (red dots) and Bremsstrahlung (blue dots)
models. The errors bars in both the spectra and residual plots are
1-$\sigma$. The insets show the $\chi^2$ plots as a function of the
power law index, $\Gamma$, and $N_{\rm H}$ (upper inset), and of the
plasma temperature, $kT$, and $N_{\rm H}$ (lower inset) of the
spectral fits to the {\it Chandra} ACIS-I spectrum of LS~I~+61\,303
using absorbed power law and Bremsstrahlung models, respectively. The
contours represent 68$\%$, 90$\%$, and 99$\%$ confidence levels.}
\label{xspec}
\end{figure}

\lsi\ is clearly detected in our {\it Chandra} observation, with
a background-subtracted averaged count rate of 1.067$\pm$0.008
cnts~s$^{-1}$ in the energy band 0.5--10.0 keV. This count rate would
have indeed resulted in considerable pile-up for a standard
observational setup. The X-ray spectrum (Figure~\ref{xspec}) shows a
broad peak at $\sim$1.6 keV and a hard energy tail that extends up to
9 keV. Assuming a foreground interstellar absorption with solar
abundances and the absorption cross-sections of \citet{BM92}, the
X-ray spectrum of \lsi\ can be reasonably well fitted by either an
absorbed power law, or an absorbed Bremsstrahlung model
(Fig.~\ref{xspec}). The best fit parameters, goodness of the fit
(reduced $\chi^2$), and implied source flux and luminosity in the
energy range of 0.5--10.0 keV for these two models are listed in
Table~\ref{tbl-xrays}. In both cases, the unabsorbed X-ray luminosity
of \lsi\ in the energy range of 0.5--10.0 keV is $\sim 10^{34}$
ergs~s$^{-1}$.

\begin{deluxetable}{lcc} 
\tablewidth{0pc}
\tablecaption{X-ray spectrum best-fit parameters \label{tbl-xrays}\tablenotemark{a}}
\tablehead{
\colhead{Model} & \colhead{Power-law} & \colhead{Bremsstrahlung}
} 
\startdata
$\chi^2$/DoF & 413.95/370=1.12 & 428.24/370=1.16\\
$N_{\rm H}$ (10$^{21}$ cm$^{-2}$) & 5.5$\pm$0.5 & 4.8$\pm$0.4 \\
$\Gamma$ or kT &  1.53$\pm$0.07  & 17$^{+6}_{-4}$ keV  \\
$f_{\rm obs}^{\rm 0.5-10.0\, keV}$ &  1.87$\pm$0.16  & 1.84$\pm$0.05 \\
$L^{\rm 0.5-10.0\, keV}$  &  1.18$\pm$0.11  &  1.11$\pm$0.04
\enddata
\tablenotetext{a}{Best fit parameters to the X-ray spectrum for
LS~I~+61\,303\, obtained using absorbed power-law  and
Bremsstrahlung models. The flux between 0.5 and 10.0 keV
is expressed in 10$^{-11}$ ergs cm$^{-2}$ s$^{-1}$ and the
luminosity in 10$^{34}$ ergs s$^{-1}$ and for an assumed distance of
2~kpc.}
\end{deluxetable}

\begin{figure}
\epsscale{1.}
\plotone{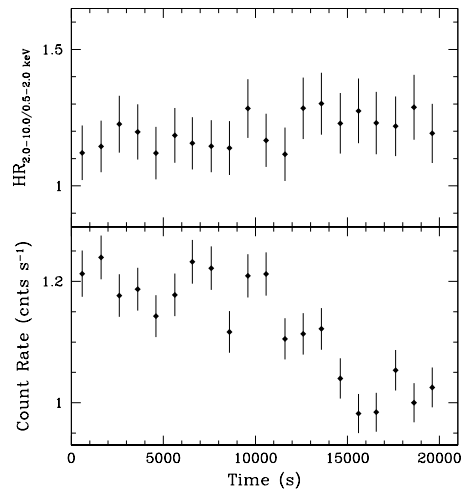}
\caption{
Temporal evolution of the {\it Chandra} ACIS-I hardness ratio (top)
and count rate in the energy band 0.5--10.0 keV (bottom). The bin size
is 1,000 s.\ and the error bars are 1-$\sigma$.}
\label{lc_hr}
\end{figure}

\begin{deluxetable*}{rlrrrrr}
\tablewidth{0pc}
\tablecaption{X-ray observations of LS~I~+61\,303 in the $\sim$1--10 keV
range\label{tab:xray-hist}\tablenotemark{a}} 
\tablehead{
\colhead{Obs.\ ID} & \colhead{Date}    & \colhead{Exposure} &
\colhead{$\phi_\textrm{orb}$ }  & \colhead{$\phi_\textrm{sup}$  } & \colhead{$\Gamma$} & 
\colhead{$f$}
} \startdata
A0 & 03-02-1994 & 18.0 & 0.20 & 0.61 &  $1.71 \pm 0.08$ & $0.72$ \\
A1 & 09-02-1994 & 19.3 & 0.43 & 0.61 &  $1.83 \pm 0.08$ & $0.52$ \\
\hline
R0 & 01-03-1996 &  8.9 & 0.79 & 0.07 &-- \quad\quad\quad& $0.84$ \\
R1 & 04-03-1996 &  9.3 & 0.91 & 0.07 &-- \quad\quad\quad& $1.21$ \\
R2 & 07-03-1996 &  9.4 & 0.03 & 0.07 &-- \quad\quad\quad& $0.69$ \\
R3 & 10-03-1996 &  9.1 & 0.11 & 0.07 &-- \quad\quad\quad& $0.65$ \\
R4 & 13-03-1996 & 10.0 & 0.22 & 0.07 &-- \quad\quad\quad& $0.96$ \\
R5 & 16-03-1996 &  9.5 & 0.33 & 0.07 &-- \quad\quad\quad& $1.32$ \\
R6 & 18-03-1996 &  9.5 & 0.42 & 0.08 &-- \quad\quad\quad& $2.00$ \\
R7 & 24-03-1996 & 11.0 & 0.64 & 0.08 &-- \quad\quad\quad& $1.22$ \\
R8 & 26-03-1996 & 13.7 & 0.71 & 0.08 &-- \quad\quad\quad& $1.03$ \\
R9 & 30-03-1996 & 14.6 & 0.87 & 0.08 &-- \quad\quad\quad& $0.92$ \\
\hline
S0 & 22-09-1997 & 12.0 & 0.31 & 0.41 & $1.68 \pm 0.16$ & $0.48 \pm 0.01$ \\
S1 & 26-09-1997 &  8.6 & 0.45 & 0.41 & $1.56 \pm 0.09$ & $1.40 \pm 0.01$ \\
\hline
X0 & 05-02-2002 &  6.4 & 0.56 & 0.37 & $1.60 \pm 0.02$ & $1.39 \pm 0.01$ \\
X1 & 10-02-2002 &  6.4 & 0.76 & 0.37 & $1.53 \pm 0.02$ & $1.35 \pm 0.01$ \\
X2 & 17-02-2002 &  6.4 & 0.01 & 0.37 & $1.74 \pm 0.05$ & $0.50 \pm 0.01$ \\
X3 & 21-02-2002 &  7.5 & 0.18 & 0.38 & $1.57 \pm 0.02$ & $1.36 \pm 0.01$ \\
X4 & 16-09-2002 &  6.4 & 0.97 & 0.50 & $1.60 \pm 0.07$ & $1.39 \pm 0.01$ \\
X5 & 27-01-2005 & 48.7 & 0.60 & 0.02 & $1.62 \pm 0.01$ & $1.29 \pm 0.01$ \\
  &            &      &      &      & $1.83 \pm 0.01$ & $0.40 \pm 0.01$ \\
\hline
C0 & 07-04-2006 & 49.9 & 0.03 & 0.28 & $1.25 \pm 0.09$ & $0.71 \pm 0.18$ \\
C1 & 25-10-2006 & 20.0 & 0.61 & 0.40 & $1.53 \pm 0.07$ & $1.87 \pm 0.02$
\enddata
\tablenotetext{a}{The first column shows an ID composed of a letter
standing for the X-ray satellite (A = {\it ASCA}~\citep{leahy97}, R =
{\it RXTE}~\citep{harrison00}, S = {\it BeppoSAX}~\citep{sidoli06}, X
= {\it XMM-Newton}~\citep{sidoli06,che06}, C = {\it
Chandra}~\citep[][and this work]{Paredes07}) and an ordering
index. Observation C1 corresponds to the one performed during our
multiwavelength campaign. The exposure is in ks. $\phi_\textrm{orb}$
and $\phi_\textrm{sup}$ are the orbital and super-orbital phases of
the beginning of the observation, computed using $T_0 = 43366.275$,
$P_\textrm{orb} = 26.4960$ days and $P_\textrm{sup} = 1667$ days
\citep{gre02}. The photon index ($\Gamma$) is obtained from a fit to
an absorbed power-law model. The flux ($f$) is expressed in 10$^{-11}$
ergs cm$^{-2}$ s$^{-1}$, and is integrated between 2-10 keV for {\it
RXTE}, {\it BeppoSAX} and {\it XMM-Newton}, between 0.3-10 keV for
observation C0 and between 0.5-10 keV for C1.}
\end{deluxetable*}

We have also investigated the short-term variability of \lsi\ X-ray
flux and hardness ratio (defined as the ratio of the count rate
between 2.0--10.0 keV over that between 0.5--2.0 keV), shown in
Figure~\ref{lc_hr}. The {\it Chandra} ACIS-I count rate of \lsi\ in
the energy band of 0.5--10.0 keV varied by a 25$\%$, from $\sim 1.25
\pm 0.03$ cnts~s$^{-1}$ down to $\sim 1.00 \pm 0.03$ cnts~s$^{-1}$,
within a timescale of 1-2 hours, showing a clear decline in the final
third of the observation. The probability for this variation to be a
statistical fluctuation of a constant flux is less than 10$^{-15}$, as
obtained from a $\chi^2$ fit of a constant function to the data
set. Fast flux variations of up to a 70$\%$ within few hours
time-scale have been reported by \cite{sidoli06}. In their case, these
were accompanied by variations in the hardness ratio (HR = 2--12
keV/0.3--2 keV) of about $30\%$, which our data do not confirm. It
must be noted, however, that assuming a linear correlation between
flux and HR, we should observe less than $\sim 10\%$ variations in the
HR during our observation, which is below our statistical error.

\begin{figure}
\epsscale{1.}
\plotone{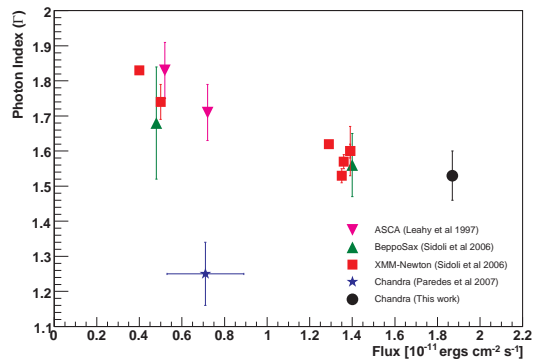}
\caption{Correlation between the X-ray flux and photon index for the
existing observations of LS~I~+61~303 in the $\sim$1--10 keV energy
band, obtained by using public data from the literature and our own
data. (See main text for details).}
\label{xray-corr}
\end{figure}

Table~\ref{tab:xray-hist} summarizes the results of the observations
of LS~I~+61\,303 in the $\sim$1--10 keV energy range existing in the
literature, including that obtained in our {\it Chandra} observations.
We measure a flux of $1.9 \times 10^{-11}$ ergs cm$^{-2}$ s$^{-1}$ and
a photon index 1.53. The historical values obtained over several years
range between $[0.4-2.0] \times 10^{-11}$ ergs cm$^{-2}$ s$^{-1}$ for
the flux and $[1.25-1.83]$ for the photon index. Therefore, we observed
the source in a particularly high and hard state. This is shown also
in Figure~\ref{xray-corr}, where we plot the historical values of the
photon index against the flux, in the soft X-ray range. The Pearson's
correlation coefficient between these two quantities is $r=-0.46$, but
becomes $r=-0.91$ when the outlier point obtained by \citet{Paredes07} is
removed from the computation. This is, to our knowledge, the first
time that this correlation is shown, and confirms the claim by
\cite{sidoli06} that the source is harder when it is brighter.

\subsection{VHE $\gamma$-ray Results}

\lsi\ was detected with MAGIC only on October 27 ($\phi_\textrm{orb} =
0.66$), with a significance of 4.5$\sigma$. For the rest of the
nights, no detection above 2$\sigma$ was found, and we derived the
corresponding upper limits to the integral flux (see
Table~\ref{tbl-log}).

On October 27, the measured average flux above 300 GeV corresponds to
15$\%$ of the Crab Nebula flux at these energies. The VHE $\gamma$-ray
source is point-like for the MAGIC angular resolution ($0.1^\circ$)
and the location is compatible with that of \lsi. The energy spectrum
is well fitted by an unbroken power-law with index $\alpha = -2.7 \pm
0.5 \pm 0.2$, where the quoted errors correspond to the statistical
and systematic uncertainties, respectively. No significant variations
of the absolute flux were detected within that night.

Previous observations of this source with the MAGIC
\citep{alb06,alb08} and VERITAS \citep{acc08} telescopes have shown a
point-like source, with a peak of $\sim 15\%$ Crab flux intensity at
orbital phase $\phi_\textrm{orb} \sim 0.65$, and a spectral index
$\alpha \sim -2.6$, hence in agreement with the results derived from
our observations.

\begin{figure}
\epsscale{1.1}
\plotone{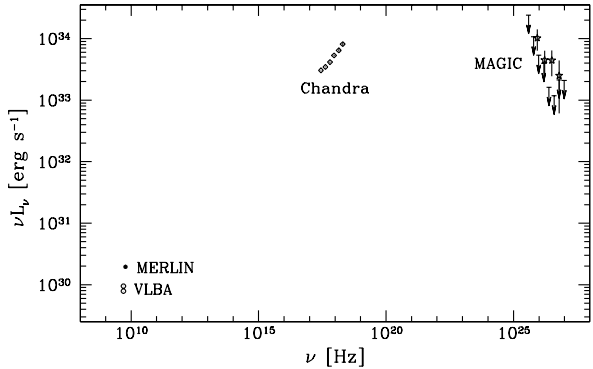}
\caption{Quasi-simultaneous LS~I~+61\,330\, spectrum
including radio (VLBA, open circles), X-rays ({\it Chandra}, squares)
and VHE $\gamma$-rays (MAGIC, stars) data from the period 25--26
October 2006, along with the data from November 2006: average MERLIN
flux density (filled circle) and upper limits from MAGIC (arrows).
\label{sed}}
\end{figure}

\section{Discussion}  %

The comparison between VHE $\gamma$-ray and radio data during the
October campaign (see Figure~\ref{fig:obs}) shows a detection at TeV
energies for $\phi_\textrm{orb} = 0.66$ with a flux level of $\sim
15\%$ of the Crab Nebula flux, during a period when the radio emission
is constant at 35 mJy. During the November campaign, however, there is
no detection at TeV energies, while the radio data show a peak flux
twice as high as in October. \citet{alb06} reported on radio and TeV
peaks detected almost simultaneously, while for our campaing we see
the TeV peak for a flat and low radio flux and a radio peak for no
significant TeV emission. Therefore, we exclude a general TeV-radio
correlation. A plausible explanation is that the emissions are
produced by different particle populations. On the other hand, the
detections at X-ray and TeV energies, both at particularly high flux
values and one day apart, might point to a correlation between X-ray
and TeV fluxes, and hence to the fact that both radiations are
produced by the same population of particles. Yet, our data are too
scarce to make any solid conclusion at present.

The results obtained by radio imaging at different angular scales show
that the size of the radio emitting region of \lsi\ is constrained
below $\sim$6~mas ($\sim$12 projected AU), and the presence of
persistent jets above this scale is therefore excluded. 

We resolve a radio-emitting region at 5.0 GHz with VLBA, extending
east-southeast from the brighter, unresolved emitting core. The
outflow velocity implied by our observations is at least of
$\sim$0.1~c, in agreement to what previously suggested by
\cite{che06}, for an over-pressured pulsar wind, although recent
hydrodynamical simulations
\citep{bogo08} show that the shocked pulsar wind could be relativistic.

The comparison of our VLBA image from 25 October with an image at 8.4
GHz obtained at a similar orbital phase but 10 orbital cycles apart
shows a high level of similarity between them. It is worth noting that
\cite{dha06} obtained high-resolution radio images of the source along
a complete orbital cycle, and found an extended feature whose
orientation with respect to the central core varied by $360^\circ$
along the orbit. Therefore, the similarity in the morphology and
orientation between both images suggests a significant level of
periodicity and stability of the physical processes involved in the
radio emission. This result points to the fact that the extended radio
emission is produced by the interaction of steady flows (from a
relativistic pulsar wind, jet and/or stellar wind) rather than by the
interaction of such an outflow with wind clumps. We note that if the
radio emission is produced by a milliarcsecond scale jet, the required
stability and periodic behavior of such a jet is difficult to
reconcile with the non-persistent nature of a large scale ($\sim$100
mas), putative relativistic jet, as deduced from our MERLIN
observations in combination with those obtained by \cite{mas04}.

Finally, we combine our X-ray data from 25 October, the TeV data from 26
October and the average VLBA in both nights to produce a
quasi-simultaneous multi-wavelength spectrum, including radio, X-ray
and VHE $\gamma$-ray observations, which we display in
Figure~\ref{sed}. However, the simultaneous data cover only a
small range of the orbital phase of \lsi. Given the high variability
of the physical conditions of the system along the orbit, more
simultaneous, multiwavelength data, and particularly involving longer
exposure times, orbital phase coverage and redundancy, will shed
further light in our understanding of this peculiar object.

\acknowledgements 

We thank the IAC for the excellent working conditions at the ORM. We
also thank the support given by the {\it Chandra} telescope and the
MERLIN interferometer through their Discretionary Director Time
programs, as well as the observing time provided by the e-EVN, and the
NRAO VLA and VLBA interferometers via Target of Opportunity
proposals. We are grateful to the anonymous referee for the useful
comments and suggestions which have improved greatly this paper. The
support of the German BMBF and MPG, the Italian INFN, the Spanish MEC,
the ETH Research Grant TH~34/04~3 and the Polish MNiI Grant 1P03D01028
is gratefully acknowledged. This research is partially supported by
the Ram\'on y Cajal programme of the Spanish MEC. The European VLBI
Network (EVN) is a joint facility of European, Chinese, South African
and other radio astronomy institutes funded by their national research
councils. e-VLBI developments in Europe are supported by EC DG-INFSO
funded Communication Network Development project, EXPReS, contract
no. 02662.  MERLIN is a National Facility operated by the University
of Manchester at Jodrell Bank Observatory on behalf of STFC. This work
has benefited from research funding from the European Community's
sixth Framework Programme under RadioNet R113CT 2003 5058187. The
National Radio Astronomy Observatory is a facility of the National
Science Foundation operated under cooperative agreement by Associated
Universities, Inc.

\end{document}